\documentclass[11pt,a4paper,fleqn]{article}   
\catcode`\@=11
\@addtoreset{equation}{section}
\def\theequation{\arabic{section}.\arabic{equation}}
\def\appendix{\renewcommand{\thesection}{\Alph{section}}\setcounter{section}{0}
              \renewcommand{\theequation}
            {\mbox{\Alph{section}.\arabic{equation}}}\setcounter{equation}{0}}

\def\maketitle{\thispagestyle{empty}\setcounter{page}0\newpage
                \renewcommand{\thefootnote}{\arabic{footnote}}
                  \setcounter{footnote}0}
\renewcommand{\thanks}[1]{\renewcommand{\thefootnote}{\fnsymbol{footnote}}
               \footnote{#1}\renewcommand{\thefootnote}{\arabic{footnote}}}

\renewcommand{\title}[1]{\begin{center}\Large\bf #1\end{center}\rm\par\bigskip}
\renewcommand{\author}[1]{\begin{center}\Large #1\end{center}}
\newcommand{\address}[1]{\begin{center}\large #1\end{center}}

\def\babs{\hrule\par\begin{description}\item{Abstract: }\it} 
\def\eabs{\par\end{description}\hrule\par\medskip\rm}
\renewcommand{\date}[1]{\par\bigskip\par\sl\hfill #1\par\medskip\par\rm}

\def\segue{\qquad\Longrightarrow\qquad} 
\def\hs{\qquad}               
\def\nn{\nonumber}            
\def\beq{\begin{eqnarray}}    
\def\be{\begin{eqnarray}}
\def\eeq{\end{eqnarray}}      
\def\ee{\end{eqnarrayn}}
\def\ap{\left.}               
\def\at{\left(}               
\def\aq{\left[}               
\def\ag{\left\{}              
\def\cp{\right.}              
\def\ct{\right)}              
\def\cq{\right]}              
\def\R{{\hbox{{\rm I}\kern-.2em\hbox{\rm R}}}}   
\def\H{{\hbox{{\rm I}\kern-.2em\hbox{\rm H}}}}   
\def\N{{\hbox{{\rm I}\kern-.2em\hbox{\rm N}}}}   
\def\C{{\ \hbox{{\rm I}\kern-.6em\hbox{\bf C}}}} 
\def\Z{{\hbox{{\rm Z}\kern-.4em\hbox{\rm Z}}}}   
\def\Tr{\mathop{\rm Tr}\nolimits}                  
\def\dir{/\kern-.7em D\,}                          
\def\al{\alpha}
\def\ga{\gamma}
\def\de{\delta}

\def\om{\omega}

\def\be{\begin{equation}}
\def\ee{\end{equation}}
\def\bea{\begin{eqnarray}}
\def\eea{\end{eqnarray}}

\def\nn{\nonumber}

\renewcommand{\title}[1]{\begin{center}\Large\bf #1\end{center}\rm\par\bigskip}
\renewcommand{\author}[1]{\begin{center}\Large #1\end{center}}

\begin{document}

\title{On the Stability of a class of Modified   Gravitational 
 Models  }
\author{ Guido Cognola\thanks{cognola@science.unitn.it}, Monica Gastaldi,
Sergio Zerbini\thanks{zerbini@science.unitn.it}}
\address{
Dipartimento di Fisica, Universit\`a di Trento \\ 
and Istituto Nazionale di Fisica Nucleare \\ 
Gruppo Collegato di Trento, Italia}

\begin{abstract}
Motivated by the dark energy issue, a minisuperspace approach to the stability 
for modified gravitational models in a four dimensional cosmological 
setting  is
investigated. Specifically, after revisiting the $f(R)$ case, $R$ being 
the Ricci curvature, we present a
stability condition around a de Sitter solution valid for modified 
gravitational models of generalized  type $F(R,G, Q)$, $G$  and $Q$ being the 
Gauss-Bonnet and quadratic Riemann   invariants respectively. A generalization to 
higher order invariants is mentioned. 

\end{abstract}


\section{Introduction}

It is well known that recently it has been found strong 
evidence for an accelerated expansion of the universe, 
apparently due to the presence of an effective positive 
cosmological constant and associated with 
this acceleration there exists the so called dark energy issue 
(see for example \cite{padmanabhan}). 
 
The modified gravity models are pure gravitational alternative 
for dark energy (for a recent review and alternative approaches 
see \cite{CST,rev3}). The main idea underlying these approaches to dark
energy puzzle  is quite simple and consists in
adding to the gravitational Einstein-Hilbert action other gravitational terms 
which may dominate the cosmological evolution
during the very early or the very late universe epochs, 
but in such a way that General Relativity remains valid at intermediate epochs
and also at non cosmological scale.

In the present paper we shall consider 
a large class of modified gravitational cosmological models 
defined in a Friedmann-Robertson-Walker (FRW) space-time and we shall focus our
attention on the stability of the de Sitter solution.
We recall that the stability issue is relevant in many 
contexts. For example, in the $\Lambda$CDM  model 
it ensures that no future singularity will be present in the solution and
within cosmological models, the stability or instability around a de Sitter 
solution is of some interest at early or later times. However, we remind that
the inclusion of a cosmological term has to confront with the well known 
cosmological constant problem, an unsolved issue so far. On the other hand,
as anticipated before, the modified gravity models may offer a quite natural 
geometrical approach in the spirit of the original Einstein idea.
 
First we shall briefly revisit the class of models based on the action
(here $\kappa^2=8\pi G_N$, $G_N$ being the Newton's constant) 
\be
\nonumber
S=\frac{1}{2\kappa^2}\int d^4x \sqrt{-g} f(R) \,,
\label{genR}
\ee
where $f(R)$ depends only on the scalar curvature $R$. We will consider pure 
gravitational theories, since we are interested in the dark energy sector,
namely the property of the de Sitter critical point. 
The inclusion of ordinary matter can be done and it will be not treat 
here, even though its inclusion  is important in reconstructing the expansion 
history of the Universe and probing the phenomenological relevance of the 
models (see for example the recent paper \cite{amendola,amendola1}, 
where the case $f(R)$ has been discussed). 

In a cosmological setting, the $f(R)$ models have been 
introduced in \cite{turner,cappozziello} and investigated in many papers 
(see \cite{rev3}). 
Also the stability of the solutions has been 
discussed in several places 
\cite{cognola05,faraoni,cognola,NO,capo,NO1,NO2,amendola,bazeia07,rador,soko}. 
To this aim, different techniques have been employed, 
including  manifestly covariant and field theoretical 
approaches, where  the gauge issue has been properly taken into account. 
All these investigations are in agreement with the following conditions
which ensures the existence and the stability of the de Sitter solution:
\be
2f_0=R_0f'_0\,,
\label{fR}
\ee
\be
1 < \frac{f'_0}{R_0f''_0}\,.
\label{sR}
\ee
where $f',f''$ are the derivatives of $f(R)$ with respect to $R$ and 
$f'_0,f''_0$ are the derivatives evaluated at the value $R=R_0$. 
The first condition,Eq.~(\ref{fR}), determines the scalar curvature  of the 
de Sitter solution, while the second one, Eq.~(\ref{sR}), gives the  condition for 
the stability around the de Sitter solution. 
In a more general quadratic theory,
the stability has been investigated in \cite{herv,topo}.  

The aim of this paper is to address the same investigation to a more general
class of modified gravity, that is the one based on the action 
\be
\nonumber
S=\frac{1}{2\kappa^2}\int d^4x \sqrt{-g} F(R,G,Q) \,,
\label{genGQ}
\ee 
which depend on an arbitrary function of the scalar 
curvature $R$ , Gauss-Bonnet invariant $G$ and the quadratic Riemann invariant 
$Q$, the relation among them being
\be
G=R^2-4P+Q\,,\hs\hs
P=R_{\mu\nu}R^{\mu\nu}\,,\hs
Q=R_{\mu\nu\rho\sigma}R^{\mu\nu\rho\sigma}\,.
\label{GBinv}
\ee
The string-inspired scalar-Gauss-Bonnet gravity case $F(R,G)$ 
has been suggested in Ref.~\cite{sasaki} as a model for
gravitational dark energy while some time ago it
has  been applied to the possible solution of the initial singularity 
problem \cite{ART}. 

The investigation of different
regimes of cosmic acceleration in such string-inspired gravity models
has been carried out in
Refs.~\cite{sasaki,fGB,Sami,Mota,Calcagni,Neupane,GB,cognola06,cognola066,sami}.
In particular, in \cite{cognola06} a first attempt to the study of 
the stability of such kind of models has been carried out
using an approach based on  quantum field theory.

The method we will use in order to study  the stability for 
$F(R,G,Q)$ models  is a classical approach, which we call minisuperspace 
approach, and   it is based on a Lagrangian formalism
 \cite{vilenkin85,capozziello02,cognolaV},  inspired by the seminal 
paper \cite{staro}, where, for the first time, quantum effects were 
considered. 

With regard to this, it is well known that one-loop and two-loop quantum 
effects induce higher derivative gravitational terms in the effective 
gravitational Lagrangian and early studies on instability for quadratic terms 
have been 
investigated in \cite{staro1}. A particular case has been recently 
studied in \cite{aco}.   

For  general quadratic models, namely with linear dependence on 
$R^2$, $P$ and $Q$, the stability has been reported in \cite{herv,topo}.  

It should be stressed that the stability studied here is the one
with respect to  homogeneous perturbations. For the $F(R)$ case, the stability
criterion for homogeneous perturbations coincides with the inhomogeneous ones
\cite{faraoni}.

The content of the paper is the following. In Section II, we revisit the 
$f(R)$ models and rederive the stability condition (\ref{sR}) and then the
same analysis is extended to the Gauss-Bonnet models in Section III and to the
general case in Section IV. 
The paper ends with the conclusions.

\section{Minisuperspace approach for $f(R)$}

As already mentioned above, we shall deal with FRW isotropic and homogeneous 
models with spatial flat metric, namely 
\be
ds^2=-N^2(t) dt^2+a^2(t)d^2 \vec{ x}\,,
\label{frw}
\ee
where $t$ is the cosmic time, $a(t)$ the  cosmological 
factor and $N(t)$ an arbitrary lapse function, which describes the gauge 
freedom associated with the reparametrization invariance of the minisuperspace 
gravitational model. For the above metric, the scalar curvature reads
\be
R=6\left(\frac{\ddot a}{a N^2}+\frac{\dot{a}^2}{a^2 N^2}-
\frac{\dot a  \dot N}{a N^3}\right)\,.  
\label{r}
\ee
As usual the 'dot' over the symbol means derivative
with respect to time, so $\dot a=\frac{d a}{dt}$ and so on.
 
If one plugs this expression in the Eq.~(\ref{genR}), one obtains 
 a higher derivative Lagrangian theory. 
In order to work with a standard  (first derivatives) 
Lagrangian system, we make use of a 
Lagrangian multiplier $\lambda$  and we write 
\cite{vilenkin85,capozziello02}
\be
S=\frac{1}{2 \kappa^2}\int d \vec x \int dt N a^3 \left[ f(R)-
\lambda \left[ R-6\left(\frac{\ddot a}{a N^2}+\frac{\dot a^2}{a^2 N^2}-
\frac{\dot a  \dot N}{a N^3}\right)  \right] \right]\,.
\ee
Making the variation with respect to $R$, one gets
\be
\lambda=\frac{df(R)}{dR}\,.
\ee
Thus, substituting this value and making an integration by part one 
arrives at the Lagrangian, which will be our starting point
\be
L(a,\dot a,R,\dot R,N)=-6\,\frac{\dot a^2 a}{N}\frac{d f(R)}{d R}
-6\,\frac{\dot a a^2 \dot R}{N}\frac{d^2 f(R)}{d^2 R}+
Na^3\aq f(R)-R\frac{d f(R)}{d R}\cq\,.
\label{l}
\ee
Now, $a$ and $R$ are  Lagrangian variables and $N$ appears as an
``einbein'' Lagrangian multiplier, reflecting the parametrization invariance 
of the action. Furthermore, the three equations of motion related to
these three variables are not independent as we will show.
In the analysis of the system we can use only two equations of motion
and we can fix $N(t)$ by the gauge choice 
(this corresponds to the choice of the cosmological time).

The first Equation of motion
$\frac{\partial L}{\partial N}=0$ reads
\be
\dot f' H+f' H^2+\frac{1}{6}(f-Rf')=0\,,
\label{E}
\ee
where for convenience we have introduced 
\be
H=\frac{\dot a}{a}\,,\hs\hs f'=\frac{d f}{d R}\,,
\ee
and we have chosen the gauge $N(t)=1$. 

The conserved quantity related to the parametrization invariance 
is the energy computed with the standard Legendre 
transformation and it is vanishing on shell due to the 
equation of motion (\ref{E}) for the einbein $N$. 
This is the energy constraint.

The other equations of motion associated with the variation of 
$R$ and $a$ are respectively
\be
\dot H+2H^2-\frac{R}{6}=0\,,
\label{h}
\ee
\be
2 \ddot f'+4\dot f' H-2H^2 f'-\frac{Rf'}{3}+f=0\,.
\label{eeR}
\ee
As we already said above, this latter equation 
is redundant. In fact, taking the derivative with respect to $t$ of 
Eq.~(\ref{E}) and making use of Eq.~(\ref{h}) a direct calculation leads to 
Eq.~(\ref{eeR}), while 
equation (\ref{h}) is equivalent to (\ref{r}).

The two equations (\ref{E}) and (\ref{h}) form a 
very simple autonomous system in the two variables $R$ and $H$, namely  
\be
\dot R=-\frac{1}{f''}\at f' H + \frac{f-Rf'}{6H}\ct\,,
\nn 
\ee
\be
\dot H=\frac{R}{6}-2 H^2 \,. 
\nn
\ee
The analysis of stability is standard and consists first in 
finding the critical points $R_0,H_0$,
which are defined by imposing $\dot R=0$ and $\dot H=0$
and then in investigating 
the stability of the related linear system around such critical 
points.
In our case we have the  solutions
\be
R_0=12H_0^2\,,
\ee
\be
2f_0-R_0f'_0=0\,,
\label{MM}
\ee
which correspond to a de Sitter critical point  
with scalar curvature determined by the above condition.

The linearized system around the de Sitter critical point reads
\beq
\at\begin{matrix}
{\delta\dot R\cr\delta\dot H\cr}
\end{matrix}\ct=M
\at\begin{matrix}
{\delta R\cr\delta H\cr}
\end{matrix}\ct\,, 
\hs\hs
M=\at\begin{matrix}
{H_0   &   -\frac{4 f_0}{R_0f_0''}\cr\frac16& -4 H_0\cr}
\end{matrix}\ct\,.
\eeq
where $M$ is the stability matrix. It is easy to show  that the stability is 
ensured by the two conditions  $\Tr M<0$ and
$\det M >0$. The first one is trivially satisfied, while the second one,
making use of Eq.~(\ref{MM})  leads to
the stability condition (\ref{sR}).

\section{Minisuperspace approach for the Gauss-Bonnet model}

In this Section we shall generalize the previous  
approach to the  modified  gravitational model defined by a Lagrangian 
density of the type $F(R,G)$. Here $G$ is the Gauss-Bonnet invariant,
which for the FWR metric reads
\be
G=\frac{24\dot a^2}{a^3N^5}
\left(\ddot a N-\dot a  \dot N \right)\,,
\label{g}
\ee
while $R$ is given by Eq.~(\ref{r}).

In order to put the Lagrangian in a standard form, now we have
to use two Lagrangian multipliers, so we start from
\bea
S&=&\frac{1}{2 \kappa^2} \int d \vec x \int dt N a^3 \left\{ F(R,G)-
\lambda \left[ R-6\left(\frac{\ddot a}{a N^2}+\frac{\dot a^2}{a^2 N^2}-
\frac{\dot a  \dot N}{a N^3}\right)  \right] \cp 
\nn\\ &&
\hs\hs -\ap\mu\left[ G-\frac{24\dot a^2}{a^3 N^5} \left( \ddot a N-
\dot a  \dot N\right)\right]\right\}\,.
\eea
Making the variation with respect to $R$ and $G$ one gets
\be
\lambda=\frac{\partial F(R,G)}{\partial R}\equiv F'_R\,,
\hs\hs \mu=\frac{\partial F(R,G)}{\partial G}\equiv F'_G\,,
\ee
and by making an integration by part one
arrives at the Lagrangian
\bea
L(a,\dot a,R,\dot R,G,\dot G,N)&=&-\frac{6\dot a^2  a F'_R}{N}
-\frac{8\dot a^3 }{N^3}\frac{d F'_G}{dt}
-\frac{6\dot a a^2}{N}\frac{d F'_R}{dt}
\nn \\ &&\hs\hs\hs
+Na^3 (F-RF_R'-GF_G')\,.
\label{lg}
\eea
Of course, the couple of equations of motion related to the variations of
$R$ and $G$ are equivalent to Eqs.~(\ref{r}) and (\ref{g}),
which lead to
\be
R=6(\dot H+2H^2)\,,\hs\hs G=4H^2(R-6H^2)\,,
\label{rg}\ee
while the equation $\frac{\partial L}{\partial N}=0$ gives
\be
24 H^3 \dot F'_G+6 H^2 F'_R+6H\dot F'_G+(F-RF'_R-GF'_G)=0\,,
\label{Egen}
\ee
where we have chosen $N(t)=1$ again and we have put $H(t)=\dot a(t)/a(t)$
as above.
The conserved quantity is always the energy computed 
with the standard Legendre transformation and it is vanishing on shell.

The equation of motion related to the variation of  $a$ is
\bea
8 H^2 \ddot F'_G&+&2 \ddot F'_R+4H\dot F'_R+16H\dot F'_G(\dot H+H^2)
\nn \\ &+&
F'_R\at 4 \dot H +6H^2 \ct+F-RF'_R-GF'_G=0\,.
\label{eeGen}
\eea
Again, this latter equation is 
a consequence of the others. In fact,  taking the derivative with respect to 
$t$ of Eq.~(\ref{Egen}) and making use of Eqs.~(\ref{rg}), one obtains 
Eq.~(\ref{eeGen}).
As a consequence, in order to deal with  a first order differential 
autonomous system, we may use Eqs.~(\ref{rg})
and (\ref{Egen}), obtaining in this way
\be
\dot R=\frac{B(R,H)}{A(R,H)}\,,
\label{e1} 
\ee
\be
\dot H=\frac{R}{6}- 2H^2 \,, 
\label{e2} 
\ee
where
\be
A(R,H)=6HF'_R+48H^3F''_{RG}+96H^5 F''_{GG}\,,
\label{A}
\ee
\be
B(R,H)=-F+(R-6H^2)(F'_R+4H^2F'_G)-8H^2(R-12H^2)^2(F''_{RG}+4H^2F''_{GG})\,.
\label{B}
\ee
The critical points are defined by $\dot R=0$ and $\dot H=0$. 
As a result
\be
R_0=12H^2_0\,,\hs\hs G_0=24 H_0^4
\ee
and from $B(R_0,H_0)=0$ it follows
\be
F_0-G_0F'_G(R_0,G_0)-\frac{R_0F'_R(R_0,G_0)}{2}=0\,.
\label{fGen}
\ee
This corresponds to a de Sitter critical point  with Gauss-Bonnet invariant 
determined by the condition (\ref{fGen}) (see ref.  \cite{cognola06}).

Since 
\be
\frac{1}{A_0}\frac{\partial B_0}{\partial R}=H_0\,,
\ee
 the linearized system around de Sitter critical point reads
\beq
\at\begin{matrix}
{\delta\dot R\cr\delta\dot H\cr}
\end{matrix}\ct=M
\at\begin{matrix}
{\delta R\cr\delta H\cr}
\end{matrix}\ct\,, 
\hs\hs
M=\at\begin{matrix}
{H_0   &   -\frac{12F'_R(R_0,G_0)}{A_0}\cr\frac16& -4 H_0\cr}
\end{matrix}\ct\,.
\eeq
where $M$ is the stability matrix and $A_0,B_0$ are quantities in (\ref{A}) and (\ref{B}) evaluated at
$R_0$ and $H_0$. 
Again, requiring $\Tr M <0$ ( again trivially satisfied) and $\det M>0$ 
 one obtains  the stability condition 
\be
1<\frac{F'_R(R_0,G_0)}{R_0\aq 
F''_{RR}(R_0,G_0)+\frac23R_0F''_{GR}(R_0,G_0)+\frac19R_0^2F''_{GG}(R_0,G_0)
\cq}\,,
\label{M}
\ee
It is easy to show that this condition, when $F(R,G)=f(R)$ reproduces 
the  relation (\ref{sR}) discussed in Section II.

It has to be noted that the same result 
can be obtained starting from Eq.~(\ref{eeGen}) and 
reducing it to a first order autonomous system
by the method discussed in \cite{amendola}, but in this case, the resulting 
stability matrix is  a $3 \times 3$ matrix. 

We conclude the Section with an example. We recall that
it is interesting to investigate Gauss-Bonnet models 
defined by
\be
F(R,G)=R+f(G)\,,
\ee
since they may be relevant from a phenomenological point of view 
\cite{cognola066,cognola06}. In these special cases 
the conditions (\ref{fGen}) and (\ref{M}) read
\be
G_0f'_0-f_0=6H_0^2\,.
\label{fG}
\ee
\be
1<\frac{9}{R_0^3f''_0}\,.
\label{sg}
\ee
As a particular example let us choose
\be
f(G)=\alpha G^\gamma\,,
\label{bo}
\ee
where $\alpha$ is a dimensional constant and $\gamma$ is a dimensionless 
parameter. Eq.~(\ref{fG}) gives
\be
2\alpha(\gamma-1)G_0^{\gamma-1/2}=\sqrt6
\ee
and this implies that $\al(\ga-1)>0$,
while Eq.~(\ref{sg}) leads to $1<1/(2\gamma)$. Then the model 
is stable (around de Sitter) 
if the arbitrary parameters $\alpha$ and $\beta$ satisfy 
both the conditions
\beq
\ag\begin{array}{ll} 
\alpha(\gamma-1)>0\\
\frac{1}{2\gamma}>1
\end{array}\cp\segue
\ag\begin{array}{ll} 
\alpha<0\\
0<\gamma<\frac12
\end{array}\cp\,.
\label{COND}\eeq
All the other possible choices for the parameters $\alpha,\beta$ 
give rise to an unstable de Sitter solution. In particular,
the model with negative parameters $\al$ and $\gamma$ is unstable.

\section{Minisuperspace approach for the $F(R,G,Q)$ case}

Now we shall generalize the approach described in the previous section 
to a Lagrangian density of the  type $F(R,G,Q)$. 
The use of $G$ instead of $P$ simplifies 
the derivation. Here $Q$ is the quadratic Riemann  invariant,
which for the FWR metric (\ref{frw}) reads
\be
Q=\frac{12}{a^2N^6}
\left(\ddot a N-\dot a  \dot N \right)^2+ \frac{12 \dot a^4}{a^4N^4}\,.
\label{q}
\ee
In order to put the Lagrangian in a standard form,  now we have
to use three Lagrangian multipliers, so we start from
\bea
S&=&\frac{1}{2 \kappa^2} \int d \vec x \int dt N a^3 \left\{ F(R,G,Q)-
\lambda \left[ R-6\left(\frac{\ddot a}{a N^2}+\frac{\dot a^2}{a^2 N^2}-
\frac{\dot a  \dot N}{a N^3}\right)  \right] \cp 
\nn\\ 
 &-&\ap\mu\left[ G-\frac{24\dot a^2}{a^3 N^5} \left( \ddot a N-
\dot a  \dot N\right)\right]-
\rho\left[Q-\left(\frac{12}{a^2N^6}
\left(\ddot a N-\dot a  \dot N \right)^2+ 
\frac{12 \dot a^4}{a^4N^4}\right)\right]
\right\}\,.
\eea
Making the variation with respect to $R$, $G$ and $Q$, one gets
\be
\lambda=\frac{\partial F}{\partial R}\equiv F'_R\,,
\hs \mu=\frac{\partial F}{\partial G}\equiv F'_G\,,
\hs \rho=\frac{\partial F}{\partial Q}\equiv F'_Q\,,
\ee
and by making an integration by part, one
arrives at the Lagrangian
\bea
L&=&-\frac{6\dot a^2  a F'_R}{N}
-\frac{8\dot a^3 }{N^3}\frac{d F'_G}{dt}
-\frac{6\dot a a^2}{N}\frac{d F'_R}{dt}+ Na^3 (F-RF'_R-GF'_G-QF'_Q)
\nn \\ &&\hs\hs
  +12 F'_Q\at\frac{\dot a^4 }{aN^3}+\frac{a\ddot a^2 }{N^3}
+\frac{a\dot a^2 \dot N^2}{N^5}- \frac{2a\dot a\ddot a \dot N}{N^4}\ct\,.
\label{lgq}
\eea
The equations of motion related to the variations of
$R$, $G$ and $Q$ are equivalent to Eqs.~\ref{r}), (\ref{g}) and (\ref{q})
respectively and read
\be
R=6(\dot H+2H^2)\,,\hs G=4H^2(R-6H^2)\,,\hs Q=12H^4+\frac{1}{3}(R-6H^2)^2\,.
\label{exrgq}\ee
Here we have put $H(t)=\dot a(t)/a(t)$ and we have choose 
the ``gauge'' $N(t)=1$. 
The equation of motion for the variable $N$ gives (in the gauge $N(t)=1$)
\bea
24 H^3 \dot F'_G +6 H^2 F'_R+6H\dot F'_R+(F-RF'_R-GF'_G-QF'_Q) 
\nn\hs && \\ 
+ 24H(\dot H+H^2)\dot F'_Q 
+12F'_Q(2H\ddot H+6\dot HH^2-\dot H^2)&=&0\,,
\label{EQgen}
\eea
while the equation of motion related to the variation of $a$
again is a consequence of the other ones and for this reason 
we do not write it.

Now making  use Eqs.~\ref{exrgq})
and (\ref{EQgen}), we arrive at the following first order differential 
autonomous system again composed only by two equations,
\be
\dot R=\frac{B(R,H)}{A(R,H)}\,,
\label{e1q} 
\ee
\be
\dot H=\frac{R}{6}- 2H^2 \,, 
\label{e2q} 
\ee
where
\beq
A(R,H)&=&4HF'_Q+6H\aq F''_{RR}+8H^3(F''_{RG}+F''_{RQ}) \cp
\nn\\&&\hs
+\ap 16H^4( F''_{GG}+2F''_{GQ}+F''_{QQ})\cq +A_1 \,,
\label{Aq}
\eeq
\beq
B(R,H)&=&-F+RF'_R+GF'_G+QF'_Q-6H^2F'_R+4H^2F'_Q(R-12H^2)+ B_1\,.
\label{Bq}
\eeq
where $A_1$  and $B_1$ are trivially vanishing  when evaluated at   
the critical points  $\dot R=0$ and $\dot H=0$. They are given by
\be
A_1=8H(R-12H^2)\at F''_{RQ}+4H^2F''_{GQ}+\frac{R}{3}F''_{QQ} \ct\,,
\ee
\be
B_1=-8H(R-12H^2)^2\at F''_{RG}+4H^2F''_{GG}-F''_{QR} \ct\,.
\ee  
As a result, one has
\be
R_0=12H^2_0\,,\hs \hs G_0=Q_0=24 H_0^4
\ee
and from $B(R_0,H_0)=0$, it follows
\be
F_0-\frac{R_0^2}{6}(F'_G+Q_0F'_Q)-\frac{R_0F'_R}{2}=0\,.
\label{fGena}
\ee
This generalizes the condition (\ref{fGen}). 

The linearized system around de Sitter critical point reads
\beq 
\at\begin{matrix}{\de\dot R\cr\de\dot H\cr}\end{matrix}\ct
=M\,\at\begin{matrix}{\de R\cr\de H\cr}\end{matrix}\ct\,,
\hs\hs M=
\at\begin{matrix}{H_0&-\frac{12 H_0(F'_R+8H_0^2F'_Q)}{A_0}\cr  
\frac16&-4 H_0\cr}\end{matrix}\ct\,,
\label{}\eeq
$A_0=A(R_0,H_))$ being the quantity in (\ref{Aq}) evaluated at
the critical point. Again the trace of the $2 \times 2$ matrix $M$ is equal to 
the $-3H_0<0$, while  from $\det M>0 $, we get the stability condition   
\be
1<\frac{F'_R+\frac23R_0F'_Q}{R_0\aq
   F''_{RR}+\frac23F'_Q
     +\frac23R_0(F''_{GR}+F''_{RQ})
      +\frac19R_0^2(F''_{GG}+2F''_{GQ}+F''_{QQ})
\cq}\,.
\label{stabilita}\ee
This is the main result of our paper.
If $F$ does not depend on $Q$, 
the above condition reduces Eq.~(\ref{M}) derived in Section III.

Let us consider an example, namely the model defined by \cite{turner1}
\beq
F(R,G, Q)=R-\mu^2\,f(R,G,Q)\,,
\label{chiba-like}\eeq
\beq
f(R,G)=\frac{1}{(\al R^2+6\beta G+6 \gamma Q)^n}\,,
\hs
f_0=f(R_0,G_0,Q_0)=\frac{1}{R_0^{2n}(\al+\beta+\gamma)^n}\,,
\eeq 
where $\mu^2,\al,\beta, \gamma$ are arbitrary dimensional and dimensionless 
parameters 
and $n$ is an arbitrary integer number. 
Now from Eqs.~(\ref{fGen}) and (\ref{M}) we respectively obtain
\beq 
\frac{R_0}{2}-(n+1)\mu^2\,f_0=0\segue f_0>0\segue (\al+\beta+\gamma)^n>0\,,
\label{ex1}\eeq
\beq 
\frac{\al+\beta+\gamma}{\om_n(\al,\beta,\gamma)}< 0\,,\hs \hs
\om_n(\al,\beta,\gamma)=2n(\al+\beta+\gamma)+2\beta+\alpha\,.
\label{ex2}\eeq
Eq.~(\ref{ex1}) ensures the existence of the de Sitter solution
while (\ref{ex2}) is necessary for the stability of such a solution. 

We immediately see that (\ref{ex2}) is not satisfied if all
the parameters $\al$, $\beta$ and $\gamma$ have the same sign, 
but it is not satisfied also in some other cases. 
For example, it is never satisfied if both
$\al+\beta+\gamma$ and $\om_n(\al,\beta,\gamma)$ have the same sign.  
The cases in which $\al+\beta+\gamma$ and $\om_n(\al,\beta,\gamma)$ have 
opposite signs require a more detailed analysis. 

As an example, we may take $\alpha=-\gamma, \gamma>0$, $\beta >0$ thus
$\alpha+\beta+\gamma=\beta>0$. Then,  as soon as  
$\beta < \frac{\gamma}{2(n+1)}$, one has a De Sitter stable solution.  

We conclude this Section observing that the method we are dealing with is 
very suitable for generalization to the case where higher curvature invariants
of thecurvature are present. An interesting  example is the third order 
invariant $Q_3$, given by 
\beq
Q_3&=&R_{\mu \nu \alpha \beta}R^{\alpha \beta \rho\sigma}
R_{\rho \sigma}^{\mu \nu}
\nn\\
&=&
24\,{\frac{\dot{N}^{3}\dot{a}^{3}}{a^{3}N^{9}}}
-72\,{\frac{\dot{N}^{2}\dot{a}^{2}\ddot{a}\,}{a^{3}N^{8}}}
+72\,{\frac{\dot{N}\dot{a}\,\ddot{a}^{2}}{a^{3}N^{7}}}
-24\,{\frac{\ddot{a}^{3}}{a^{3}N^{6}}}
-24\,{\frac{\dot{a}^{6}}{a^{6}N^{6}}}\,.
\label{Q3}
\eeq 
We recall that this is the invariant which governs the two-loop 
ultraviolet divergences in pure gravity \cite{sagno,van}. 

If now we consider the special case $F(R,G,Q,Q_3)=f(R,G,Q)-bQ_3$, then 
a straightforward calculation leads to the 
following relations for the existence and the stability of 
the de Sitter solution:
\be
F_0-\frac{R_0^2}{6}(F'_G+F'_Q)-\frac{R_0F'_R}{2}-\frac{b}{24}R^3_0=0\,,
\label{fGenab3}
\ee
\beq
1<\frac{F'_R+\frac23R_0F'_Q+\frac{b}{4}R_0^2}{R_0\aq 
  F''_{RR}+\frac23F'_Q
    +\frac23R_0(F''_{GR}+F''_{RQ})
      +\frac19R_0^2(F''_{GG}+2F''_{GQ}+F''_{QQ})
 \cq
 +\frac{b}{3}R_0^2}
\,.\nn\\
\label{stabilita3}
\eeq
As an explicit example of this type, we consider the function
\be
F=R-2\Lambda_0 +a_1R^2+a_2P+a_3Q-b\,Q_3 \,,
\ee
where $P$ is given by Eq.~(\ref{GBinv}) and in order to use 
Eq.~(\ref{fGenab3} it has to be 
replaced in terms of $R,G$ and $Q$. From (\ref{fGenab3}), we get
\be
R_0-4 \Lambda_0-\frac{b}{36}R_0^3=0\,,
\label{v}
\ee
while condition  (\ref{stabilita3}) leads to
\be
\frac{12-bR_0^2}{(3a_1+a_2+a_3)R_0+\frac{b}{2}R_0^2}>0\,.
\ee
For $\Lambda$ small, one has real positive solutions.
When $b=0$ (the absence of the cubic term), the stability condition
is  
\be
a_2+a_3+3a_1>0\,,
\ee
which is the result obtained in \cite{topo}. In particular, when $a_2=0$ and
$a_3=0$, one 
obtains the well known result that models of the type $R-2\Lambda_0+a_1R^2$ 
are stable as soon as $a_1 >0$. 

As a final remark, one should note that for 
$b>0$, besides the  flat solution
one can also have a de Sitter solution induced by quantum effects \cite{staro}
starting with $\Lambda_0=0$ in the Lagrangian, the scalar curvature being
$R_0^2=36/b$.
As it is well known, and evident from Eq.~(\ref{v}), this is not true for 
the quadratic case, which requires a non vanishing cosmological constant
term $\Lambda_0$  in the initial Lagrangian. In this case, the stability 
condition is satisfied when
\be
(a_2+a_3+3a_1)<-3\sqrt{b}\,.
\ee

\section{Conclusions}

In this paper we have presented the Lagrangian minisuperspace approach to
the stability issue  around a de Sitter critical point for a class of 
modified gravitational models  depending 
on the Ricci scalar,Gauss-Bonnet and Quadratic Riemann  invariants. 
It should be stressed that the stability studied here is the one
with respect to  homogeneous perturbations. For the $F(R)$ case, the stability
criterion for homogeneous perturbations coincides with the inhomogeneous ones
\cite{faraoni}.  

With respect to other approaches, 
the method here presented has the advantage that the 
autonomous system one is dealing with, as far as the stability of  de Sitter 
solution is concerned, consists {\em always} in two 
first order differential equations, 
simplifying considerably the stability analysis for {\em arbitrary} dependence 
on the chosen curvature invariants.

The method has been first applied to the class 
of models depending only on the  Ricci scalar $R$ and 
the well known stability condition for this case has been recovered. 
Then the same  approach has been applied to the 
scalar-Gauss-Bonnet models and its generalizations and a new  general 
condition 
for the stability around a de Sitter solution has been found.  

With regard to this last issue, we would like to note that the conditions 
(\ref{fGenab3})
(existence of the de Sitter solution) and  and (\ref{stabilita3}) 
(stability condition around the de Sitter solution)  are  very general 
and they can be applied to a large  class of Lagrangians, as shown 
in the last example of Sec, IV, where a third order invariant in the 
Riemann tensor has been investigated.

As far as  the comparison with other methods is concerned, in  
reference \cite{aco}, a criterion of stability
 for a class of generalized modified gravitational models around de Sitter 
solution and described by a function of the kind 
$W(R, Q-4P)$ has been derived with a field theoretical approach. Since 
$G=R^2-4P+Q$, our stability condition can easily be related to the one 
obtained within a theoretical approach.  

Furthermore, we also have to mention Ref.~\cite{chiba}, where the model of 
Ref. \cite{turner1}  have been discussed. 
The model considered in that paper is described by the
choice
\beq 
F(R,P,Q)=R-\frac{\mu^2}{(aR^2+bP+cQ)^n}\,,
\label{chiba}\eeq
with $a,b,c$ positive parameters and $n$ a positive integer.
The instability of the model is due to the appearance of 
a spin-2 ghost. As a result, an  instability of the vacuum arises. 
In our notation, this model corresponds to the first model (\ref{chiba-like}) 
we have studied in Section IV and we have
\beq
\alpha=a+\frac{b}{4}\,, \hs\beta=-\frac{b}{4}\,\hs \gamma=c+\frac{b}{4}\,.
\eeq
When $a,b,c$ and $n$ are positive, our condition (\ref{stabilita}) is not 
satisfied and, as a consequence, the model is unstable, in agreement with 
the conclusions reported in Ref. \cite{chiba}. Thus, the instability is a 
generic feature of this class of modified models. Within a theoretical 
approach, this has been already noted in Ref. \cite{nunez}. 

Finally, we observe that the inclusion of the matter for a generic  $F$ 
might be  non trivial, but, within our approach, can be done, for example, 
along the line of Refs. \cite{amendola,amendola1}.

\section*{Acknowledgments}
We would like to thanks S.D.~Odintsov for comments and suggestions.


\begin{thebibliography}{99}


\bibitem{padmanabhan}T.~Padmanabhan, Phys.\ Rept.\ {\bf 380} 235 (2003); 
[ArXiv:astro-ph/0603114]. 

\bibitem{CST}E.~Copeland, M.~Sami and S.~Tsujikawa, [ArXiv:hep-th/0603057]. 


\bibitem{rev3}S.~Nojiri and S.~D.~Odintsov, 

Int. J. Geom. Meth. Mod. Phys. {\bf 4} 115 (2007). 
[ArXiv:hep-th/0601213]. 



\bibitem{amendola}L. Amendola, R. Gannouji, D. Polarski and S. Tsujikawa, 
  Phys. Rev. {\bf D75}083504 (2007). 
 [ArXiv: gr-qc/0612180]. 

\bibitem{amendola1}L. Amendola and S. Tsujikawa,
{\em Phantom crossing, equation-of-state singularities, and local gravity constraints in f(R) models } [ArXiv: 0705.0396]. 
 

\bibitem{turner}S. M. Carroll, V. Duvvuri, M. Trodden and M. Turner, Phys. Rev. {\bf D70}043528(2004). [arXiv: astro-ph/0306438]. 

\bibitem{cappozziello}S. Capozziello, S. Carloni, A. Troisi, {\em Quintessence without scalar fields } [ArXive: astro-ph/0303041]; S. Capozziello, S. Carloni, A. Troisi Phys. Rev. {\bf D71}043503 (2005). 

\bibitem{cognola05}G. Cognola, E. Elizalde, S. Nojiri, S. D. Odintsov and S. Zerbini, JCAP {\bf 0502 }010 (2005). [ArXiv:hep-th/0501096]. 

\bibitem{faraoni}V.~Faraoni, Ann.\ Phys.\ {\bf 317}, 366 (2005) [arXiv:gr-qc/0502015]; V. Faraoni, Phys. Rev. {\bf D 72}, 061501 (2005), 
V. Faraoni, {\em de Sitter space and the equivalence between f(R) and scalar-tensor gravity }, arXiv: gr-qc/0703044. 

\bibitem{cognola}G.~Cognola and S.~Zerbini, J.\ Phys.\ A {\bf 39} (2006) 6245 [arXiv:hep-th/0511233]. 

\bibitem{NO}S. Nojiri and S. D. Odintsov, Phys. Rev. {\bf D 68}123512 (2003). [ArXiv:hep-th/0307288]. 

\bibitem{capo}S. Capozziello, S. Nojiri, S. D. Odintsov and A. Troisi, Phys. Letts. {\bf B 639} 135 (2006). [arXiv:astro-ph/0604431]. 

\bibitem{NO1}S. Nojiri and S. D. Odintsov, Phys. Rev. {\bf D 74}086006 (2006). 

\bibitem{NO2}S. Nojiri and S. D. Odintsov, {\em Modified gravity and its reconstruction from the universe expansion history} [ ArXiv:hep-th/0611071]. 

\bibitem{bazeia07}D. Bazeia, B.Carneiro da Cunha, R. Menezes, A.Yu. Petrov {\em Perturbative aspects and conformal solutions of F(R) gravity} [ArXiv:hep-th/0701106]. 

\bibitem{rador} T. Randor,
{\em Acceleration of the Universe via f(R) Gravities and The Stability of 
Extra Dimensions} [ArXiv:hep-th/0701267]; T. Randor,
{\em f(R) Gravities a la Brans-Dicke} [ArXiv:hep-th/0702081]. 



\bibitem{soko} L. M. Sokolowski,
{\em Physical interpretation and viability of various metric nonlinear gravity theories applied to cosmology} [ArXiv:gr-qc/0702097]. 


\bibitem{herv}
J. D. Barrow and S. Hervik,
Phys. Rev. {\bf D 74}124017  (2006). 
[ArXiv:gr-qc/0610013]. 


\bibitem{topo}A.V. Toporensky, P.V. Tretyakov, {\em De Sitter stability in quadratic gravity } [ArXiv:gr-qc/0611068]. 


\bibitem{sasaki}S.~Nojiri, S.~D.~Odintsov and M.~Sasaki, Phys.\ Rev.\ D {\bf 71}, 123509 (2005) [arXiv:hep-th/0504052]. 


\bibitem{ART}I.~Antoniadis, J.~Rizos, K.~Tamvakis, Nucl.\ Phys.\ B {\bf 415}, 497 (1994) [arXiv:hep-th/9305025]; N.~E.~Mavromatos and J.~Rizos, Phys.\ Rev.\ D {\bf 62}, 124004 (2000); Int.\ J.\ Mod.\ Phys.\ A {\bf 18}, 57 (2003) 


\bibitem{fGB}S.~Nojiri, S.~D. Odintsov, Phys.\ Lett.\ B {\bf 631} 1 (2005) [arXiv:hep-th/0508049]. 

\bibitem{Sami}M.~Sami, A.~Toporensky, P.~V.~Tretjakov and S.~Tsujikawa, Phys.\ Lett.\ B {\bf 619}, 193 (2005) [arXiv:hep-th/0504154]; S.~Tsujikawa and M.~Sami, arXiv:hep-th/0608178; 

\bibitem{Mota}T.~Koivisto and D.~F.~Mota, 
Phys. Lett. {\bf B644}104 (2007), [ArXiv:astro-ph/0606078]; 
Phys. Rev. {\bf D75}023518 (2007) [ArXiv:hep-th/0609155]; Z.~Guo, N.~Ohta and S.~Tsujikawa, arXiv:hep-th/0610336; G. Calcagni, B. Carlos and A. De Felice, Nucl. Phys. B {\bf 752} 404 (2006); 

\bibitem{Calcagni}G.~Calcagni, S.~Tsujikawa and M.~Sami, Class.\ Quant.\ Grav.\ {\bf 22}, 3977 (2005) [arXiv:hep-th/0505193]; 
A.~Sanyal, Phys. Lett. {\bf B645} 1-5 (2007), [ArXiv:astro-ph/0608104]. 

\bibitem{Neupane}I.~P.~Neupane, Class. Quant. Grav. {\bf 23}7493 (2006),
 [ArXiv:hep-th/0602097]; 
B.~M.~N.~Carter and I.~P.~Neupane, JCAP {\bf 0606}004 (2006), [ArXiv:hep-th/0512262]. 

\bibitem{GB}S.~Nojiri, S.~D.~Odintsov and O.~G.~Gorbunova, J.\ Phys.\ A {\bf 39} 6627(2006), [arXiv:hep-th/0510183]; 
I.~Brevik and J.~Quiroga,  {\em Vanishing Cosmological Constant in Modified Gauss-Bonnet Gravity with Conformal Anomaly}, [ArXiv:gr-qc/0610044]. 

\bibitem{cognola06}G. Cognola, E. Elizalde, S, Nojiri, S. D. Odintsov 
and S Zerbini, Phys. Rev. {\bf D75}086002 (2007).  [ArXiv:hep-th/0611198]. 

\bibitem{cognola066}G. Cognola, E. Elizalde, S, Nojiri, S. D. Odintsov and S. Zerbini, Phys. Rev. {\bf D 73} 084007 (2006). 

\bibitem{sami}S.~Nojiri, S.~D.~Odintsov and M.~Sami, Phys.~Rev.~D {\bf 74} (2006) 046004, arXiv:hep-th/0605039. 

\bibitem{vilenkin85}A. Vilenkin, {\sl Phys. Rev.} {\bf D 32} (1885) 2511. 

\bibitem{capozziello02}S. Capozziello, Int. J. Mod. Phys. {\bf D11}4483 (2002). 

\bibitem{cognolaV}G. Cognola, S. Zerbini, Vestnik TSPU, {\bf 7}, 69 (2004). 

\bibitem{staro}A. Starobinsky, {\sl Phys.Lett.} {\bf B91} (1980) 99.   

\bibitem{staro1}A. Starobinsky and H. J. Schmidt,
 {\sl Class. Quantum Grav.} {\bf 4} 695 (1987); V. Muller, H. J. Schmidt 
and A. Starobinsky, {\sl Phys.Lett.} {\bf B 202} 198 (1988);    
H. J. Schmidt, {\sl Class. Quantum Grav.} {\bf 5} 233 (1988)


\bibitem{aco}
I. Navarro and K. Van Acoleyen,
JCAP {\bf 0603} 008 (2006), 
arXiv:gr-qc/0511045.


\bibitem{turner1}
S.M. Carroll, A. De Felice, V. Duvvuri, D. A. Easson, M. Trodden  and 
M.S. Turner,
Phys. Rev.  {\bf D 71} 063513, (2005), 
[ArXiv:astro-ph/0410031].


\bibitem{sagno}M. H.Goroff and  A. Sagnotti,
 {\sl Nucl. Phys.} {\bf B 266}709 (1986).   


\bibitem{van} A.E.M.  van den Ven,
 {\sl Nucl. Phys.} {\bf B 378}309 (1992).   


\bibitem{chiba}
T. Chiba,
JCAP {\bf 0503} 008 (2005), 
arXiv:gr-qc/0502070.

\bibitem{nunez}
A. Nunez and S. Solganik,
Phys. Letters {\bf B 608} 189 (2005), 
arXiv:hep-th/0411102.


\end{thebibliography}
\end{document}